\RequirePackage{fix-cm}
\documentclass[smallcondensed]{svjour3}       
\smartqed  
\usepackage{subfigure}
\usepackage{graphicx}
\usepackage{bm}
\usepackage{amsmath}
\usepackage{cite}
\usepackage{algorithm}
\usepackage{algorithmic}
\usepackage{setspace}
\usepackage{upgreek}
\usepackage{xcolor}
\usepackage{geometry}
\begin{document}

\title{Quantum algorithm for neural network enhanced multi-class parallel classification
}


\author{Anqi Zhang$^{1}$         \and
        Xiaoyun He$^{1}$ \and Shengmei Zhao$^{1,2^*}$
}

\institute{
1 Institute of Signal Processing Transmission, Nanjing University of Posts and Telecommunications (NUPT), Nanjing, 210003,  China \\
2 Key Lab of Broadband Wireless Communication and Sensor Network Technology, Ministry of Education, Nanjing, 210003, China \\
* E-mail:zhaosm@njupt.edu.cn \\
}

\date{Accepted: 2021.10.24} 


\maketitle

\begin{abstract}
Using the properties of quantum superposition, we propose a quantum classification algorithm to efficiently perform multi-class classification tasks, where the training data are loaded into parameterized operators which are applied to the basis of the quantum state in quantum circuit composed by \emph{sample register} and \emph{label register}, and the parameters of quantum gates are optimized by a hybrid quantum-classical method, which is composed of a trainable quantum circuit and a gradient-based classical optimizer. After several quantum-to-class repetitions, the quantum state is  optimal that the state in \emph{sample register} is the same as that in \emph{label register}. 
For a classification task of $L$-class, the analysis shows that the space and time complexity of the quantum circuit are $O(L*logL)$ and $O(logL)$, respectively. The numerical simulation results of 2-class task and 5-class task show that the proposed algorithm has a higher classification accuracy, faster convergence and higher expression ability. The classification accuracy and the speed of converging can also be improved by increasing the number times of applying multi-qubit controlled operators on the quantum circuit, especially for multiple classes classification.

\end{abstract}

\section{Introduction}
\label{intro}
Classification is one of the main problems in \emph{Machine Learning}\cite{SenPC,MinaeeS}. Based on quantum parallel processing, the related quantum algorithm is expected to exponentially speed up \cite{HarrowAW,AntonioM,ScottA}. There currently exist several kinds of quantum classifiers, one are inspired by their corresponding classical classifiers with their kernel parts replaced by quantum circuits \cite{BiamonteJ,HoWK,ErikT,LloydS,RebentrostP}, some are inspired by neural networks \cite{FarhiE,MariA,WangSQ,NathanK,Gily}, in which a plenty of qubits and quantum gates are commonly supplied to achieve the data storage\cite{FarhiE,WangSQ,NathanK} and parameter optimization\cite{Gily}, and others \cite{WiebeN,FingerhuthM,OtterbachJ,GrantE,TemmeK,AdhikaryS,ChalumuriA,ChenSYC,BhatiaAS,Adri,SoumikA} are proposed with a hybrid quantum-classical (HQC) structure where the evaluations are performed by quantum hardware while the parameters are optimized  with classical methods in classical computer.

HQC-based quantum classifiers have become a promising candidate in the Noisy Intermediate Scale Quantum(NISQ) era. For example, Otterbach \emph{et al.} \cite{OtterbachJ} investigated a hybrid quantum algorithm for unsupervised learning task, where quantum approximate optimization algorithm and gradient-free Bayesian optimization were used to train the parameters. Grant \emph{et al.}\cite{GrantE} introduced a new method of classification based on a hierarchical structure for quantum circuits for a binary classification task. Temme \emph{et al.}\cite{TemmeK} also suggested a quantum SVM classifier based on quantum variation circuit. To address the large dimensions data loading problem in quantum classifier, Adhikary \emph{et al.}  \cite{AdhikaryS} proposed a quantum classifier under the supervised learning scheme by using a strategy where all the input feature vectors were encoded in a single quNit(a N-level quantum system). After studying a hybrid classifier based on tensor network, Chen \emph{et al.}\cite{ChenSYC} further proposed an end-to-end hybrid classical-quantum classifier, in which a matrix product state (MPS)\cite{BhatiaAS} was used as a feature extractor to produce a low dimensional feature vector. Additionally, to obtain higher expression ability, Ref.\cite{Adri} presented a multiple data re-uploading method, in which a quantum circuit was organized as a series of data re-uploading and single-qubit processing units. And Ref.\cite{SoumikA} built a  model that could simultaneously manipulate two training samples. However, the algorithm for more than two training samples in qubit system has not been discussed yet.

To manipulate multiple training samples with a higher expression ability, we propose a quantum algorithm for multi-class classification task in this paper. In the algorithm, two registers are prepared, named \emph{sample register} and \emph{label register}. The training data are loaded into parameterized operators which work on the qubits of \emph{sample register} initialized as $|0\rangle$ at beginning. The basis states in \emph{label register} are used to distinguish different classes and as a controller of the parameterized operators. To obtain the states in \emph{sample register} the same as its label states, a fidelity-based cost function is adopted after the measurement of the quantum circuit, and the parameters are optimized in a classical optimizer. After several repetitions, the parameters are optimal when the cost function is converged. The classification information can be detected by measurements on \emph{sample register} after the testing data is input to the trained quantum circuit.

Different from only 2-class were manipulated simultaneously in Ref.\cite{SoumikA}, we train multi-class data, more than 2-class, simultaneously in the qubit system. In addition, we use multiple qubit to express the labels of classification while only one qubit was used in Ref.\cite{SoumikA}. Furthermore, the parameterized operators of $i$-th class in our proposed algorithm are applied multiple times to the qubits to increase the expression ability of the quantum circuit. 

This paper is arranged as follows: First, the details of the quantum classification algorithm are given in Sec. 2. Next, time and space complexity of the quantum part in the proposed algorithm are analyzed in Sec. 3. Then, two experiments of classification are performed to testify the proposed algorithm in Sec. 4. Finally, some conclusions are summarized in Sec. 5.

\section{Quantum algorithm for neural network enhanced multi-class parallel classification}
\label{sec:2}
\begin{figure}[!htp]
	\centering
	\includegraphics[width=0.9\textwidth]{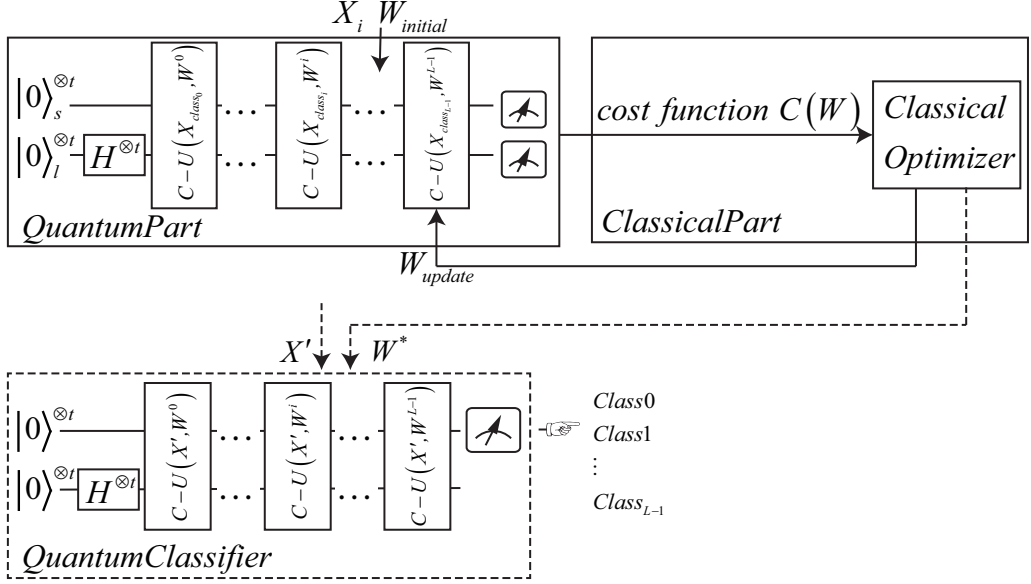}
	\caption{The framework of the proposed quantum classifier algorithm. Arrows in the figure represent the flow of information. In the training processes (body lines), the training data and the parameters are fed into the quantum circuit. The cost function $C(W)$ is obtained after measuring both \emph{sample register} and \emph{label register} and minimized by a classical optimizer. $W_{update}$ is then re-input to the quantum circuit for next data-loading. When the cost function converges after several iterations, the optimal parameters $W^{*}$ are obtained. In the testing processes (dot lines), the unclassified data $X^{'}$ is put into the quantum circuit with the optimal parameters, and the classification information of $X^{'}$ can be obtained by measuring the \emph{sample register}.}
	\label{Fig:1}
\end{figure}
In this section, we present the proposed quantum algorithm for neural network enhanced multi-class parallel classification in details.

Let's define a training dataset $S=\{X_{class_i},f(X_{class_i})\}^{N}_{i=1}$, where $X_{class_i} =(x^{i}_{1},\ldots,x^{i}_{n})$ is a sample data with $n$ dimensions, $f(X_{class_i})=0,\ldots,L-1$ is the class label corresponding to $X_{class_i}$. $M$ training data are randomly selected  from each class for optimizing the parameters, $N=M\times L$.

The framework for the proposed algorithm is given in Fig.\ref{Fig:1}. In the training processes, the initial qubits in \emph{label register} are transformed into the computational basis states by Hadamard operators, and the training data for different classes are loaded into the quantum circuit. Then, the measurements on two registers are performed for the calculation of the cost function. The parameters in cost function are optimized by a gradient-based optimization algorithm in a classical computer and re-input to the quantum circuit for next data-loading. After several iterations, the parameters are optimal when the cost function is converged. In the testing processes, the unclassified data and the optimal parameters are fed into the quantum circuit, and the classification information of the unclassified data can be obtained when the measurement results on the \emph{sample register} is performed. We present the detail of each part in the following.

\subsection{Data loading}
\label{sec:2.1}
\begin{figure}[!htp]
	\centering
	\includegraphics[width=0.75\textwidth]{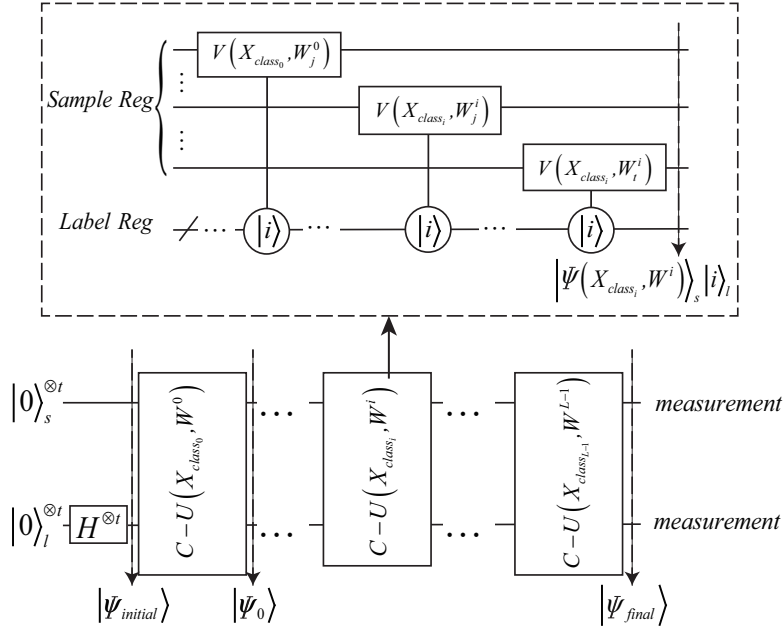}
	\caption{The details of the quantum circuit in the proposed algorithm. According to the basis state in \emph{label register}, the training data are loaded into the multi-qubit controlled operators $C$-$U(X_{class_{i}},W^{i})$(the squares), where $i=0,\cdots,L-1$.  $C$-$U(X_{class_{i}},W^{i})$ can further described by the dot square. Here, $V(X_{class_{i}},w^{i}_{j})$ is an operator on the $j$-th qubit of \emph{sample register} when the basis state in \emph{label register} is $|i\rangle$. The $i$-th multi-qubit controlled operator is formed by $t$ $|i\rangle$-controlled operators.}
	\label{Fig:2}
\end{figure}

The detailed structure of the quantum circuit is shown in Fig.\ref{Fig:2}. There are two registers, the \emph{sample register} and the \emph{label register}. Both registers have $t$ qubits which is depended on $L$, that is, $t=\lceil\log_{2}L\rceil$, where $\lceil.\rceil$ is the ceiling function. Each qubit in the two registers is initialized as $|0\rangle$. Thus, the initial quantum state after Walsh-Hadamard operators is,

\begin{equation} \label{eqn1}
|\Psi_{initial}\rangle=|0\rangle^{\otimes t}_{s}H^{\otimes t}|0\rangle^{\otimes t}_{l}=(\frac{1}{\sqrt{2^{t}}}\sum_{i=0}^{2^{t}-1}|0\rangle^{\otimes t}_{s}|i\rangle_{l}).
\end{equation}
where $i=0,1,\ldots,2^{t}-1$, $|i\rangle$ is the $i$-th computational basis state in the \emph{label register}, the subscripts $s,l$ denote the sample and the label register. For simplicity, we call the states in the two registers as the sample state and the label state, respectively. It is shown that $|\Psi_{initial}\rangle$ is a superposition state, whose component contains the sample state $|0\rangle^{\otimes t}_{s}$ and the label state $|i\rangle_{l}$.

In Fig.\ref{Fig:2}, the multi-qubit controlled operator, $C\verb|-|U(X_{class_{i}},W^{i})$, is used to upload the training data $X_{class_{i}}$ and all parameters $W^{i}$ of the $i$-th class into the quantum circuit, where the label state is the controlling state, and the sample state is the controlled state. Firstly, with the parameters, the training data of $i$-th class are fed into the parameterized operators $V(X_{class_i},w^{i}_{j})$, where $w^{i}_{j}$, $j=1,\dots,t$, indicates the parameters of $i$-th class working on the $j$-th qubit and $W^{i}=\{w^{i}_{1},...,w^{i}_{t}\}$.

A data-encoding strategy is adopted to load data into quantum circuit in the proposed algorithm. The $n$ dimensions vectors of training data $X_{class_i}$ and its parameter $w^{i}_{j}$ are divided into $K$ parts at first, where $K= n/3 $. For each part, the data are introduced by using a rotation of one qubit, that is, $SU(2)\sim U(\varphi_1,\varphi_2,\varphi_3)$ \cite{Adri}, where $\varphi_1,\varphi_2,\varphi_3$ are angle arguments to code $x^k_{1},x^k_{2}, x^k_{3}$ respectively. Moreover, each data point can be uploaded with weight  $w^{k}_{j},j=1,2,3$. 
All weights will play a similar role as weights in artificial neural networks. Altogether, each layer gate can be taken as $U(w^{k}\circ x^{k})$. Therefore, the controlled $V(X_{class_{i}},w^{i}_{j})$ is formed by successively using $SU(2)$ $K$ times,
\begin{equation} \label{eqn12}
\begin{split}
V(X_{class_i},w^{i}_{j}) = U(w^{K}\circ x^{K})\cdots  U(w^{k}\circ x^{k})\cdots U(w^{1}\circ x^{1}),
\end{split}
\end{equation}
where $w^k\circ x^k=(w^k_{1}x^k_{1},w^k_{2}x^k_{2},w^k_{3}x^k_{3})$. If $n$ can not divided by 3, $n$ is enlarged by supplementing the data with 0 elements.

Then, controlled by $|i\rangle$, the parameterized operators $V(X_{class_{i}},w^{i}_{j})$ are applied to each qubit in \emph{sample register.} The transformations of each qubit are,

\begin{equation} \label{eqn2}
\begin{split}
V(X_{class_i},w^{i}_{1})|0\rangle_{s}=|s_{1}\rangle_{s}, \\
\vdots      \\
V(X_{class_i},w^{i}_{j})|0\rangle_{s}=|s_{j}\rangle_{s}, \\
\vdots       \\
V(X_{class_i},w^{i}_{t})|0\rangle_{s}=|s_{t}\rangle_{s},
\end{split}
\end{equation}
where $j=1,\ldots,t,s_{j}\in\{0,1\}$. All qubits in \emph{sample register} and \emph{label register} form the state 
$|\Psi(X_{class_i},W^{i}) \rangle_s|i\rangle_l$ in Fig.\ref{Fig:2}.

As shown in Fig.\ref{Fig:2}, after applying $C\verb|-|U(X_{class_{0}},W^{0})$, the quantum state is,

\begin{eqnarray}\label{eqn3}
|\Psi_0\rangle &=&\frac{1}{\sqrt{2^{t}}}(C\verb|-|U(X_{class_0},W^{0})|0\rangle^{\otimes t}_{s}|0\rangle_{l}+\sum_{i=1}^{2^{t}-1}|0\rangle^{\otimes t}_{s}|i\rangle_{l})\\ \nonumber
&=&\frac{1}{\sqrt{2^{t}}}(|\Psi(X_{class_0},W^{0})\rangle_{s}|0\rangle_{l}+\sum_{i=1}^{2^{t}-1}|0\rangle^{\otimes t}_{s}|i\rangle_{l}).
\end{eqnarray}

Thus, the training data for $class_{0}$ is loaded into the superposition where the label state is $|0\rangle_l$. Next, by applying $C\verb|-|U(X_{class_i},W^{i}), i=1,\cdots,L-1$ on the superposition state sequentially, the final state should be,

\begin{eqnarray}\label{eqn4}
|\Psi_{final}(W)\rangle &=&\frac{1}{\sqrt{2^{t}}}(C\verb|-|U(X_{class_0},W^{0})|0\rangle^{\otimes t}_{s}|0\rangle_{l}+...+ \\ \nonumber
& &C\verb|-|U(X_{class_i},W^{i})|0\rangle^{\otimes t}_{s}|i\rangle_{l}+...+ \\ \nonumber
& &C\verb|-|U(X_{class_{L-1}},W^{L-1})|0\rangle^{\otimes t}_{s}|L-1\rangle_{l})\\ \nonumber
&=&\frac{1}{\sqrt{2^{t}}}(\sum_{i=0}^{2^{t}-1}|\Psi(X_{class_i},W^{i})\rangle_{s}|i\rangle_{l}) \\ \nonumber
&=&\frac{1}{\sqrt{2^{t}}}(\sum_{i=0}^{L-1}|\Psi(X_{class_i},W^{i})\rangle_{s}|i\rangle_{l})
.
\end{eqnarray}

Here, $L$ is  an integer power of 2. Otherwise, 
Eq.(\ref{eqn4}) can be rewritten as,

\begin{eqnarray}\label{eqn5}
|\Psi_{final}(W)\rangle &=&\frac{1}{\sqrt{2^{t}}}(C\verb|-|U(X_{class_0},W^{0})|0\rangle^{\otimes t}_{s}|0\rangle_{l}+...+ \\ \nonumber
& &C\verb|-|U(X_{class_i},W^{i})|0\rangle^{\otimes t}_{s}|i\rangle_{l}+...+ \\ \nonumber
& &C\verb|-|U(X_{class_{L-1}},W^{L-1})|0\rangle^{\otimes t}_{s}|L-1\rangle_{l} +\sum_{i=L}^{2^{t}-1}|0\rangle_{s}|i\rangle_{l})\\ \nonumber
&=&\frac{1}{\sqrt{2^{t}}}(\sum_{i=0}^{L-1}|\Psi(X_{class_i},W^{i})\rangle_{s}|i\rangle_{l}+\sum_{i=L}^{2^{t}-1}|0\rangle_{s}|i\rangle_{l}).
\end{eqnarray}

The objective of the training process is to make each sample state $|\Psi(X_{class_i},W^{i})\rangle_{s}$ equals to its label state $|i\rangle_{l}$, that is,
\begin{equation} \label{eqn6}
|\Psi(X_{class_i},W^{i})\rangle_{s} \rightarrow |i\rangle{s}.
\end{equation}

If one time of the multi-qubit controlled operators can not realize the transformation of Eq.(\ref{eqn6}), we can apply multiple times of the multi-qubit controlled operators to finish it, that is
\begin{equation} \label{eqn7}
C\verb|-|U(X_{class_i},W^{i_m})\ldots C\verb|-|U(X_{class_i},W^{i_1})|0\rangle_{s}^{\otimes t}|i\rangle_{l}=|i\rangle_{s}|i\rangle_{l},
\end{equation}
where $m$ indicates the number of times the multi-qubit controlled operators are used in quantum circuit for loading training data of each class. Each multi-qubit controlled operator in Eq.(\ref{eqn7}) has different parameters but with the same training data of $class_i$.

\subsection{Cost function and parameters optimization}
\label{sec:2.2}
We adopt the fidelity\cite{KerstinB} between the final output state of quantum circuit and the optimal output averaged over the training data as the cost function. Therefore, the cost function is given as,
\begin{equation} \label{eqn8}
C(W)=\frac{1}{M}\sum_{r=1}^{M}(1-|\langle\Psi_{optimal}|\Psi^{r}_{final}(W)\rangle|^{2}),
\end{equation}

\begin{equation} \label{eqn9}
|\Psi_{optimal}\rangle=\frac{1}{\sqrt{2^{t}}}(\sum_{i=0}^{2^{t}-1}|i\rangle_{s}|i\rangle_{l}),
\end{equation}
where $M$ is the number of samples used for training processes.

Finally, the measurement operator is designed as,
\begin{equation} \label{eqn10}
O = |\Psi_{optimal}\rangle\langle\Psi_{optimal}|.
\end{equation}

Here, we minimize the cost function Eq.(\ref{eqn8}) in a classical optimizer until the final state,
$|\Psi_{final}\rangle$, is closer to the optimal state, $|\Psi_{optimal}\rangle$. After several quantum-classical iterations, the parameters are trained as the optimal ones, and  the state $|\Psi(X_{class_i},W^{i})\rangle_{s}$  is closer to its  label state $|i\rangle_{s}$.

\subsection{Quantum classifier}
\label{sec:2.3}
When the optimal parameters $W^{*}$ are used in quantum circuit, the quantum circuit in  Fig.\ref{Fig:1} becomes a quantum classifier. That is, when the unclassified data $X'$ is fed into the quantum circuit, the classification information can be obtained by the measurements performed on the \emph{sample register}. A collection $\{|i\rangle\langle i|,i=0,1,...,L-1\}$ of measurement operators is used in this process. The measurement result $|i\rangle$ with the highest probability which corresponds to the $i(=0,1,...,L-1)$ is the classification information of $X'$. 

Note that if $L$ is not an integer power of 2, the quantum state obtained at the end of the quantum classifier is,

\begin{equation}\label{eqn11}
|\Psi(X')\rangle=\frac{1}{\sqrt{2^{t}}}(\sum_{i=1}^{L}|\Psi(X',W^{*})\rangle_{s}|i\rangle_{l}+\sum_{i'=0,i'\neq i}^{2^{t}-1}|0\rangle_{s}|i'\rangle_{l}),
\end{equation}
which is shown that the probability of the measurement result of $|0\rangle$ is at least $1-L/2^{t}$. Hence, 0 is mistakenly considered as the label of $X'$. Therefore, in the case of Eq.(\ref{eqn11}), the basis state of $|0...0\rangle_{s}|0...0\rangle_{l}$ is not selected to loading data in the training processes. We can choose the measurement result with the second highest probability as the label of $X'$.



\section{Complexity analysis}
\label{sec:3}
In this section, we analyze the space and time complexity of quantum circuit in the proposed algorithm. Here, the time cost and space cost are defined as the number of quantum gates and the number of qubits required to execute the circuit, respectively.

\begin{lemma} 
	For a $L$-class classification task ($L$ is assumed the integer power of 2),
	$2^{t}t(km+\frac{5}{2})+\frac{13}{4}2^{t}-2t+12$ quantum gates are required in the proposed quantum classification algorithm.
\end{lemma}
\textbf{Prove}:
We prove the \pmb{lemma 1} by operator decomposition. After the superposition generated by using $t$ Walsh-Hadamard gates in quantum circuit, $2^{t}\times m$ multi-qubit controlled operator $C\verb|-|U(X_{class_i},W^{i})$ are required for loading training data according to Eq.(\ref{eqn4}) and Eq.(\ref{eqn7}), and $t$ $|i\rangle$-controlled operators are needed in each $C\verb|-|U(X_{class_i},W^{i})$ as shown in Fig.2.

One $|i\rangle$-controlled operators can be decomposed into Toffoli gates, $X$ gates and one-qubit controlled gates, according to the discussion in [28]. If the controlling state is $|1...1\rangle$, $|i\rangle$-controlled operators can be implemented by Toffoli gates and the controlled $V(X_{class_i},w^{i}_{j})$ operator. Otherwise,  the controlling state should be converted to $|1...1\rangle$ by $X$ gates before applying Toffoli gates.

Firstly, we calculate the number of Toffoli gates needed in $|i\rangle$-controlled operator. $|i\rangle$-controlled operator is controlled by $t$ qubits, and a $t$-qubit $|i\rangle$-controlled operator need $t-1$ Toffoli gates as mentioned in Ref.\cite{NielsenMA}. Other $t-1$ $|i\rangle$-controlled operators are applied with the same way as shown in Fig.\ref{Fig:2}. For the reason that these $|i\rangle$-controlled operators are controlled by the same computational basis state $|i\rangle$, so the Toffoli gates in the two adjacent $|i\rangle$-controlled operators can be removed. Therefore, only $2(t-1)$ Toffoli gates are required for $C\verb|-|U(X_{class_i},W^{i})$ totally. Note that the processes of loading data into the quantum circuit ($m$ times) can be achieved by repeating the application of controlled $V(X_{class_i},w^{i}_{j})$ here, therefore, the number of Toffoli gates has no relationship with $m$.

Next, the number of $X$ gates is the same as the number of $|0\rangle$ in the basis states because of the requirement of flipping $|0\rangle$ to $|1\rangle$ before applying Toffoli gates. So, $2^{t}\times t$ $X$ gates are used in the whole quantum circuit. However, two adjoined $X$  gates are equivalent to one unit gate $I$, which in practice saves the number of used $X$ gates.  Thus, $X$ gates required in this algorithm are $2^{(t-2)}(2t+5)-2t+2$.  The same as the Toffoli gate, the number of $X$ gates is not related to $m$.

Finally, $k$ one-qubit controlled gates are used to form controlled $V(X_{class_i},w^{i}_{j})$ when loading data into quantum circuit as Eq.(2) and  $2^{t}\times t\times k\times m$ one-qubit controlled gates gates are required totally.

Therefore, the number of quantum gates used by the quantum circuit is $2^{t}t(km+\frac{5}{2})+\frac{13}{4}2^{t}-2t+12$.

\begin{lemma}
	For a $L$-class classification task, the proposed quantum classification algorithm requires $3t-1$ qubits totally.
\end{lemma}

\textbf{Prove}: With a classification task for $L$-class, the number of qubits in \emph{sample register} and \emph{label register} are both $t$. As described in \pmb{lemma 1}, $t-1$ auxiliary qubits are needed to form $|i\rangle$-controlled operators in the quantum circuit. The total number of qubits needed in the quantum circuit is,
\begin{equation}
t-1 + 2*t = 3t-1,
\end{equation}
where $t=\lceil\log_{2}L\rceil$. 

In summary, the time and space complexity of quantum circuit in the proposed algorithm are $O(L*logL)$ and $O(logL)$, respectively.

\section{Simulation}
\label{sec:4}
In this section, we demonstrate the feasibility of the proposed quantum classifier algorithm. The samples of training and testing data come from the MNIST data set (handwritten digit number). We present two examples here, one is a two-class classification task for digit numbers 1 and 7, and the other is a five-class classification task for digit number 1, 2, 4, 7 and 9. Here, 2000 training samples and 500 testing samples for each class are selected randomly from the MNIST data set. Every sample (the handwritten digit number image) is processed to 32 dimensions by Rough Grid Feature method\cite{PaoloC}. The quantum circuit of the algorithm is performed by using PennyLane\cite{VilleB} module in Python and the cost function is trained using Adam optimizer in a classical computer.

\begin{figure}[htbp]
	\centering
	\subfigure[]{
		\includegraphics[width=0.45\textwidth]{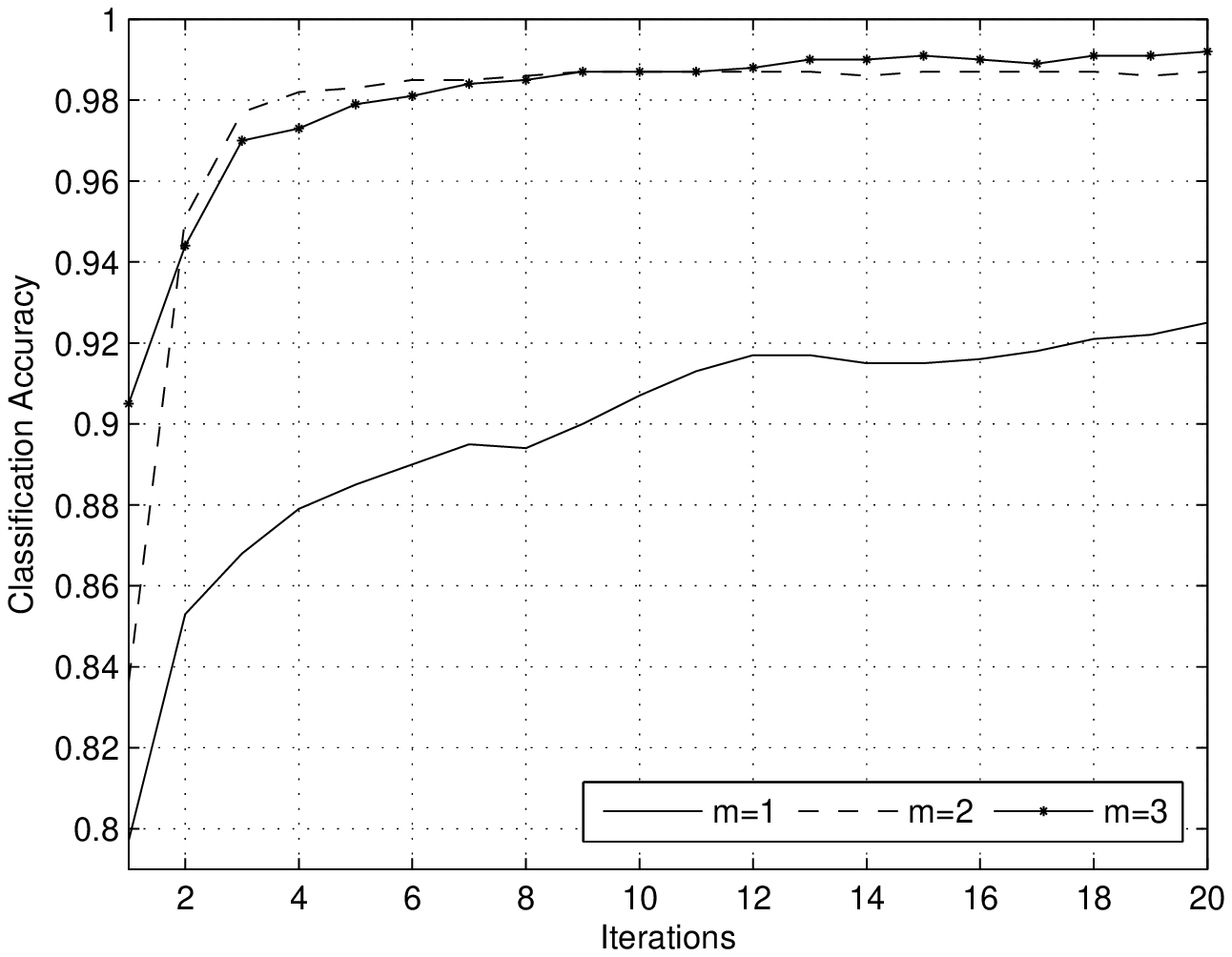}
	}
	\subfigure[]{
		\includegraphics[width=0.45\textwidth]{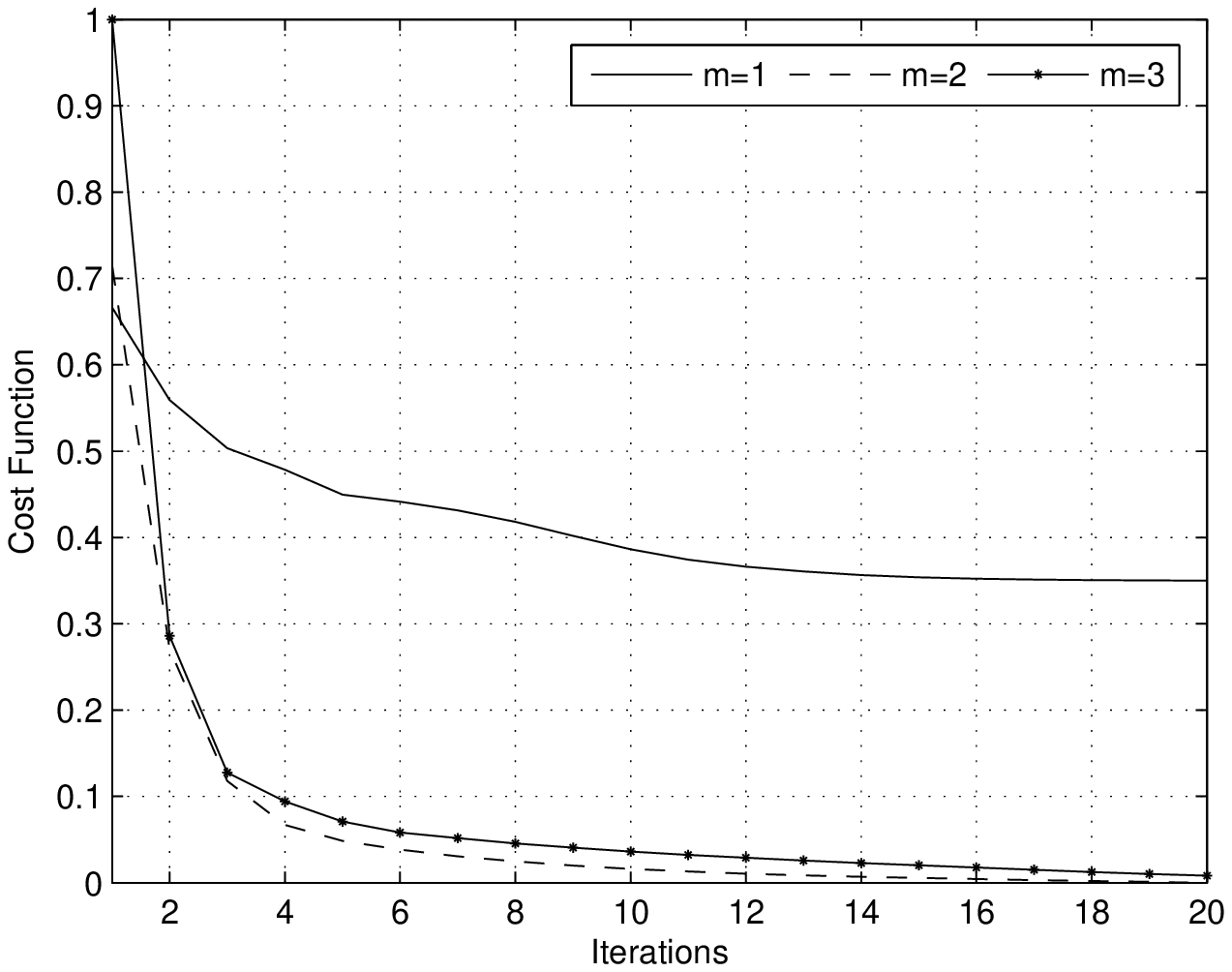}
	}
	\caption{The classification accuracy and the cost function against the iterations for the two-class classification task, where $m$, the number times of the multi-qubit controlled operators for each class in quantum circuit, is setup to 1,2, or 3. (a) is the classification accuracy and (b) is the cost function.}
	\label{Fig:8}
\end{figure}

Fig.\ref{Fig:8} shows the classification accuracy and the cost function against the iterations for the two-class classification task, where Fig.\ref{Fig:8}(a) is for the accuracy and Fig.\ref{Fig:8}(b) is for the cost function. Here, $m$, the number times of the multi-qubit controlled operators used in quantum circuit, is setup to 1, 2, and 3 respectively. The results in Fig.\ref{Fig:8}(a) show that the proposed quantum classifier algorithm  can reach a higher classification accuracy for the similar digit numbers 1 and 7. Compared to $m=1$, the classificaiton accuracy is highly improved when $m=2$ and $m=3$. The results in Fig.\ref{Fig:8}(b) show that the proposed algorithm has a fast convergences, especially, there is only 4 iterations where $m=2$ and $m=3$. It is indicated that the multiple application of the multi-qubit controlled operators can increase the expression ability of the quantum circuit. But, for the two-class classification task, it is enough for $m=2$  since the results from $m=2$  and $m=3$ are almost the same. 

\begin{figure}[htbp]
	\centering
	\subfigure[]{
		\includegraphics[width=0.45\textwidth]{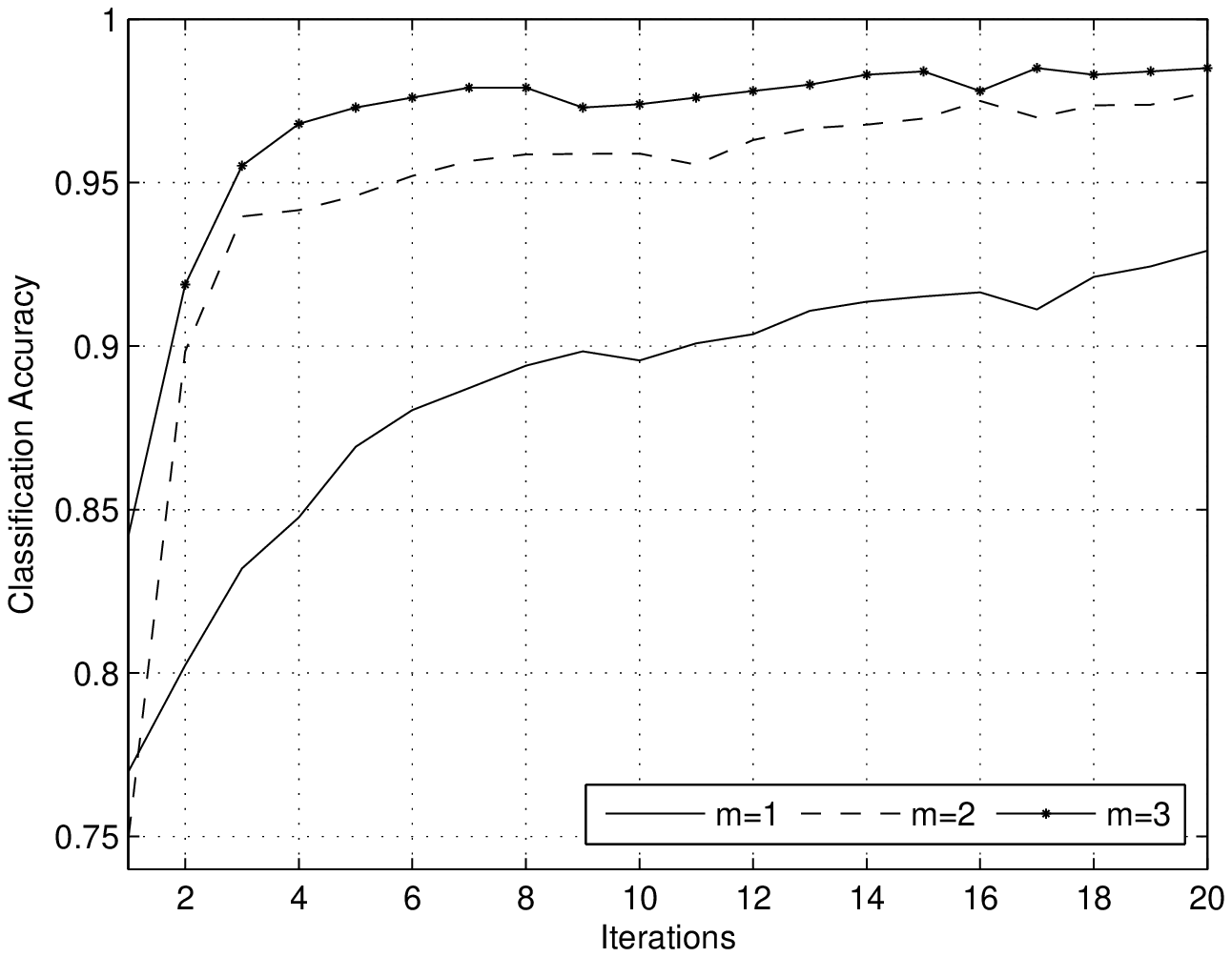}
	}
	\subfigure[]{
		\includegraphics[width=0.45\textwidth]{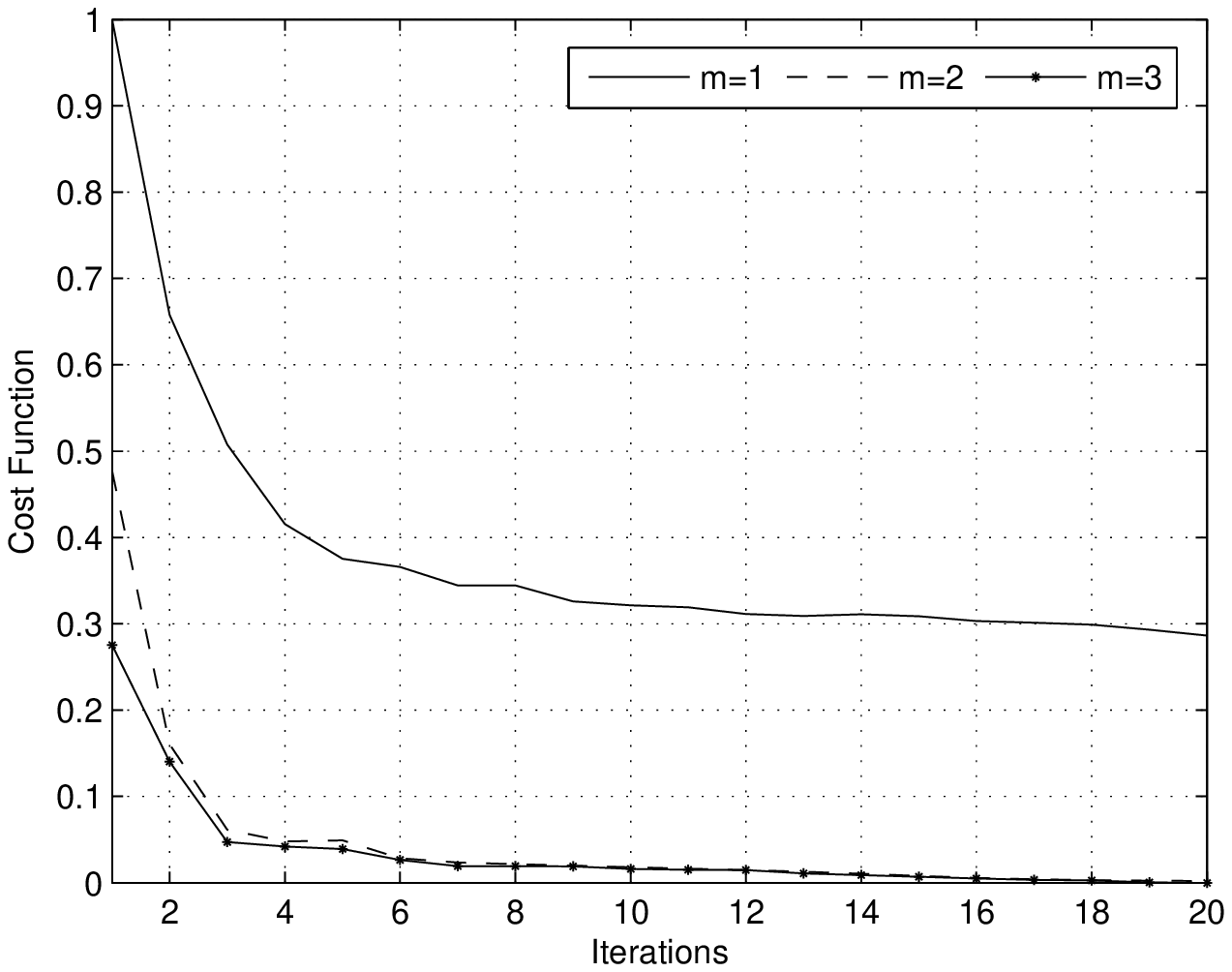}
	}
	\caption{The classification accuracy and the cost function against iterations for the five-class classification task, where $m$, the number times of the multi-qubit controlled operators for each class in quantum circuit, is setup to 1,2, or 3. (a) is the classification accuracy and (b) is the cost function.}
	\label{Fig:9}
\end{figure}

Fig.\ref{Fig:9} shows the classification accuracy and the cost function against iterations for the five-class classification task, where Fig.\ref{Fig:9}(a) is for the the accuracy and Fig.\ref{Fig:9}(b) is for the cost function, and $m$, the number times of the multi-qubit controlled operators used in quantum circuit, is setup to 1,2, and 3 respectively. Fig.\ref{Fig:9}(a) shows that the proposed algorithm can reach a higher classification accuracy for both the similar hand-writing digits(1 and 7) and difficult recognized writing digits(2, 4 and 9) and the classification accuracy improves as $m$ increased. Fig.\ref{Fig:9}(b) shows that the proposed algorithm for the multi-class task also has a faster convergence in 3 iterations when $m$=2 and $m$=3. It is indicated that the expression ability of the quantum circuit of multi-class classification task can obviously increase as the number times of the multi-qubit controlled operators used.

\section{Conclusion}
\label{sec:5}
In the paper, we have proposed a quantum algorithm for multiple classification in a hybrid classical-quantum circuit. By applying Toffoli gates, $X$-gates and one-qubit controlled gates, we have achieved to load the training data into the quantum circuit by using the multi-qubit controlled operators for multiple class simultaneously. Additionally, a fidelity-based cost function has been adopted in the corresponding classical optimizer to obtain the optimal parameters. The analysis results have shown that the time and space complexity of the quantum circuit  are $O(L*logL)$ and $O(logL)$, respectively. Additionally, the simulation results have shown that the proposed quantum algorithm has have a higher classification accuracy for multiple classes with a higher expression ability. The classification accuracy and the speed of converging have been improved by the increasing of the number times of applying multi-qubit controlled operators on the quantum circuit, especially for multiple (greater than two) classification tasks.

\begin{acknowledgements}
This work is supported by the National Natural Science Foundation of China (61871234), and Postgraduate Research $\& $ Practice Innovation Program of Jiangsu Province   (Grant KYCX19\_0900).
\end{acknowledgements}




%
%


\end{document}